\documentstyle[prb,aps,epsf,,twocolumn]{revtex} 

\newcommand{\be}{\begin{equation}}              
\newcommand{\ee}{\end{equation}}                
\newcommand{\bea}{\begin{eqnarray}}             
\newcommand{\eea}{\end{eqnarray}}
\newcommand{\nn}{\nonumber}
\newcommand{\bm}[1]{\mbox{\boldmath${#1}$}}
\newcommand{\rr}{{\bf r}}
\newcommand{\qq}{{\bf q}}

\begin{document}
\flushbottom

\draft
\wideabs{		

\title{Hypernetted-chain study of broken rotational symmetry states
	for the $\bm{\nu}$ = 1/3 fractional quantum Hall effect and
	other fractionally filled Landau levels}

\author{Orion Ciftja and C.\ Wexler}

\address{Department of Physics and Astronomy, 
         University of Missouri--Columbia, 
         Columbia, Missouri 65211}

\date{\today}

\maketitle

\begin{abstract}
We investigate broken rotational symmetry (BRS) states for the
fractional quantum Hall effect (FQHE) at 1/3-filling of the valence
Landau level (LL).  Recent Monte Carlo calculations by Musaelian and
Joynt {[}J.\ Phys.: Condens.\ Matter {\bf 8}, L105 (1996){]} suggest
that Laughlin's state becomes unstable to a BRS state  for some
critical finite thickness value.  We study in detail the properties of
such state by performing a hypernetted-chain calculation that gives
results in the thermodynamic limit, complementing other methods which
are limited to a finite number of particles.  Our results indicate
that while Laughlin's state is stable in the lowest LL, in higher LLs
a BRS instability occurs, perhaps indicating the absence of FQHE at
partial fillings of higher LLs.   Possible connections to the newly
discovered liquid crystalline phases in higher LLs are also
discussed. 
\end{abstract}
%
\pacs{PACS: 
	73.43.-f,	
	73.43.Lp, 	
	73.43.Nq, 	
	73.20.Mf,	
	64.70.Md.	
}
\vspace{0.27cm}
}                      


\section{Introduction}
\label{sec:intro}
\vspace{-0.3cm}


Recently, a pl{\ae}thora of new phenomena has emerged in the
transitional regions between different plateaus of the Hall
conductance \cite{lilly99a,du99,shayegan99,eisenstein00b1} 
of  Landau levels (LL) with index $L \ge 2$.
Near half-filling of the valence LL, extreme anisotropy has
been measured in the magnetotransport below temperatures ca.\ 100 mK,
\cite{lilly99a,du99,shayegan99} accompanied by smooth
non-linearities.\cite{lilly99a}  In addition, reentrant integer
quantum Hall effect (RIQHE) regions with striking breakdown features
and new phase transitions (presumed to be quantum in origin
\cite{fradkin99}) have been seen near 1/4 filling  of the valence LL.
\cite{cooper99}

The anisotropic behavior has been attributed to the formation of a
nematic phase of the two dimensional electron system (2DES) which
undergo a nematic to isotropic transition at higher
temperatures. \cite{fradkin99,cw2000} 
Similarly, the reentrant regions are believed to be ``bubble phases''
similar to Wigner crystals, but with several electrons per bubble, or
possibly new electronic hexatic states. \cite{fradkin99}

The motivation of our work is to study these numerous liquid
crystalline phases present in partially filled LLs by means of
many-body trial wavefunctions with broken rotation symmetry (BRS),
similar to those proposed by Musaelian and Joynt (MJ)  \cite{joynt} in
the context of the fractional quantum Hall effect (FQHE) (which are
essentially generalizations of Laughlin's wavefunction
\cite{laughlin}): 
\bea
\label{Psialpha}
  	\Psi_{\alpha}(z_1, \ldots ,z_N) &=& \\
	&& \hspace{-2.5cm} \prod_{i>j}^{N} \,
		(z_i-z_j) (z_i-z_j-\alpha) (z_i-z_j+\alpha) 
		\; e^{- \frac{1}{4}\sum_{i=i}^{N} |z_i|^2} \nn
	\,, 
\eea
where $z_j = x_j + i \, y_j$ is the complex 2D coordinate of $j$-th
electron and $\alpha$ is a complex number (we work in units of
the magnetic length: $l_0^2=\hbar/eB=1$).  This wavefunction
represents a homogeneous liquid state with filling factor $\nu = 1/3$,
lies entirely in the lowest LL (LLL),  and for $\alpha \neq 0$ has 
nematic order (for $\alpha = 0$ we recover Laughlin's wavefunction,
which is obviously isotropic). 

This wavefunction represents, therefore, a good starting point to
consider nematic QH systems (even though the filling factor is
``incorrect'' for the newly discovered anisotropic states, see below
for further analysis), by facilitating the systematic study the energy
dependence of BRS states for diverse  physical parameters (LL index,
width of the 2DES, etc).  In addition, the possibility of BRS states
for $\nu = 1/3$ is intriguing by itself, since for $\alpha \neq 0$
highly damped low-energy modes exist, \cite{damping} strongly
modifying the dynamics and possibly suppressing the FQHE. 

In Ref.\ \onlinecite{joynt} MJ investigated the possibility
of BRS in a FQHE system using Eq.\
(\ref{Psialpha}) and performing Monte Carlo (MC) simulations in a disc 
geometry.  Their results suggest that the Laughlin fluid becomes
unstable towards BRS states in quantum wells whose thickness exceeds a 
critical value depending on the electronic density. 

In this work we study the BRS state for 1/3-filling of the valence LL
(i.e.\ $\nu = M+1/3$ with $M$ integer) by using the hypernetted-chain
(HNC) method. \cite{ripka,caillol,hansen,zabolitzky,chakraborty}    
This method allows us to compute physical quantities in the
thermodynamic limit, without the limitations of using a finite number
of particles that hinder other techniques, where the extrapolation to
the thermodynamic limit is not totally unambiguous.  We find that,
contrary to MJ's results, the Laughlin state is stable in the LLL,
whereas a BRS instability is possible in higher LLs (perhaps
indicating, analogously to the arguments of
Ref. \onlinecite{koulakov}, why there is no ordinary FQHE at, e.g.\
$\nu = 7/3$).

In Sec.\ \ref{sec:hnc_theory} we present the basic theoretical
calculations needed to determine the stability of an isotropic or BRS
state.  A detailed description of the HNC formalism in the context of 
the BRS wave function is given in Sec.\ \ref{sec:hnc_method_BRS}.
The results for the BRS state in the LLL and their extension in higher
Landau levels are discussed in Sec.\ \ref{sec:results}.  Finally, in
Sec.\ \ref{sec:conclusions} we discuss our results, and analyze how
they can be extended to more realistic filling factors.

\vspace{-0.3cm}
\section{Basic theory}
\label{sec:hnc_theory}
\vspace{-0.3cm}

In this work we propose to study the stability of different states by
using trial wavefunctions like Eq.\ (\ref{Psialpha}).  We are
interested, therefore, to calculate the energy in each of these states 
to find the optimum value for the sole free parameter $\alpha$.  

We first consider the situation in the LLL, namely the state with
$\nu=1/3$.  Since the BRS wave function is completely in the LLL the
kinetic energy per particle is quenched at the lowest cyclotron energy 
\begin{equation}
\frac{1}{N} 
\frac{\langle \Psi_{\alpha}|\hat{K}
|\Psi_{\alpha}\rangle}{\langle \Psi_{\alpha}|\Psi_{\alpha} \rangle}=
 \frac{1}{2} \hbar \omega_{c} \, ,
\label{kinetic}
\end{equation}
where $\omega_c = eB/m$ is the cyclotron frequency.  The potential, or 
correlation, energy per electron is:
\be
\label{potential}
	E_{\alpha} = \frac{1}{N} 
	\frac{\langle \Psi_{\alpha}|\hat{V}| \Psi_{\alpha}\rangle}
	{\langle \Psi_{\alpha}|\Psi_{\alpha} \rangle} 
= 	\frac{\rho}{2} \int d^2r \,  V(r) \,  \left[g(\rr) - 1\right] \,,
\ee
where $V$ represents the electron-electron, electron-background, and
background-background interaction; and $g(\rr)$ is the
(angle-dependent) pair distribution function given by 
\be
\label{eq:gr}
g(\rr) = \frac{N(N-1)}{\rho^2} 
        \frac{ \int d^2r_3 \cdots d^2r_{N} 
                | \Psi_\alpha (\rr_1 \cdots \rr_{N})|^2 }
        { \int d^2r_1 \cdots d^2r_{N} 
                | \Psi_\alpha (\rr_1 \cdots \rr_{N})|^2 } \,,
\ee
where $\rr = \rr_2-\rr_1$.  The following {\em sum rule} can be
easilty proven $\rho \int d^2r \left[ g(\rr)-1 \right] = -1$, 
and is a convenient check for numerical procedures.
For an ideal 2D sample the interaction is a pure Coulomb potential
$V(r) \simeq e^2/\epsilon r$ , while in samples with finite thickness a
reasonable choice is the Zhang Das Sarma (ZDS) potential \cite{ZDS} 
$V(r)=e^2/\epsilon \sqrt{r^2+\lambda^2}$, where $\lambda$ is of the
order of the sample thickness.  Alternatively, the correlation energy
can be computed in reciprocal space: 
\be
\label{eq:ener_sq}
E_\alpha = \frac{1}{2} \int \frac{d^2q}{(2\pi)^2} \, \tilde{V}(q) \, 
\left[ S(\qq) - 1 \right] \,,
\ee
where $\tilde{V}(q)$ is the 2D Fourier transform \cite{ft} (FT) of
$V(r)$ and $S(\qq)$ is the static structure factor:
\be
\label{eq:sq}
S(\qq) - 1 =  \rho \, {\rm FT}[g(r) - 1] \,.
\ee

While both $g(\rr)$ and $S(\qq)$ are angle-dependent (e.g.\ see Figs.\
\ref{fig:gr} and \ref{fig:sq}), because the interaction potential is
centrally symmetric, the energy $E_\alpha$ depends only on the 
angle-averaged pair distribution function or static structure factor
defined as:  
\be
\label{angleav}
\overline{g}(r) = \int_{0}^{2 \pi} \frac{d \theta}{2 \pi} \; g(\rr) \,, 
\hspace{0.5cm}
\overline{S}(q) =  \int_{0}^{2 \pi} \frac{d \theta_{q}}{2 \pi} \; S(\qq) \,.
\ee

The determination of either the pair distribution function or the
structure factor is generally a complicated integral problem that
needs to be solved for each LL.  However, its is known that if
transitions to other LLs are neglected (i.e.\ a {\em single-LL
approximation}), $g(\rr)$ and $S(\qq)$ at higher LL are simply related 
to those at the LLL ($L=0$) by means of a convolution or product
respectively.  We will apply this approximation (which, moreover,
quenches the kinetic energy in higher LLs as well).  It is then,
sufficient to compute these distribution functions once in the LLL and 
then the correlation energy per electron is given by
\be
\label{eq:ener_sq_N}
E^L_\alpha = \frac{1}{2} \int \frac{d^2q}{(2\pi)^2} \, 
        \tilde{V}_{\rm eff}(q) \, \left[ S(\qq) - 1 \right] \,,
\ee
where $\tilde{V}_{\rm eff}(q) \equiv \tilde{V}(q) \, [L_L (q^2/2)]^2$. 
$L_L(z)$ are Laguerre polynomials, and $S(\qq)$ is calculated in the
LLL ($L=0$).

In what follows we compute $g(\rr)$ and $S(\qq)$ in the LLL using the
HNC method.

\vspace{-0.3cm}
\section{The HNC method for the broken rotational symmetry state}
\label{sec:hnc_method_BRS}
\vspace{-0.3cm}

Integral equation techniques such as the HNC theory \cite{ripka,caillol}
allow an accurate evaluation of the pair distribution function
and related quantities associated with a Jastrow wave function.
In particular they are extremely useful for calculations that
are performed in the thermodynamic limit.
They have been widely used in the study of classical~\cite{hansen}
and quantum fluids.~\cite{chakraborty,zabolitzky}
However the HNC method for the BRS wave function is a slightly different
from that for the Laughlin wave function, since correlations and 
related quantities  depend on both distance and relative
angle between a pair of particles.

The main quantity to be calculated in a HNC expansion is the
pair distribution function $g(\rr)$ [Eq.\ (\ref{eq:gr})], or 
equivalently the structure factor $S(\qq)$ [Eq.\ (\ref{eq:sq})].
These may then be used in conjunction with Eqs.\ (\ref{potential}),
(\ref{eq:ener_sq}), or (\ref{eq:ener_sq_N}) to determine the energy per 
electron for arbitrary values of the BRS parameter $\alpha$ or the 2D
width $\lambda$.

Although the BRS wave function is a Fermi wave function, its modulus
square,
\begin{equation}
\label{modulus}
	|\Psi_{\alpha}(z_{1}, \ldots ,z_{N})|^2=
	e^{\sum_{i>j}^{N} u(z_i - z_j)} \;
	e^{-\sum_{i=1}^{N} \frac{|z_i|^2}{2}} \,,
\end{equation}
where $u(z)=\ln |z|^2+ \ln |z-\alpha|^2+ \ln |z+\alpha|^2$,
can be viewed as a symmetric Jastrow wave function with pair
correlations and single-particle terms.  Therefore it is possible to
apply the Bose HNC formalism. \cite{fhnc}  In 
order to compute Eq.\ (\ref{eq:gr}) one needs some small parameter in
which to expand perturbatively (and re-sum a subset of diagrams).  For
standard systems like Bose liquid $^4$He, the pair correlation is
short-range and heals to $1$ for large distances, therefore the
function $\exp[u(r_{ij})]-1$ provides a possible expansion parameter
[note that in order to apply the Bose HNC expansion, the correlation 
(pseudo) potential has to satisfy the conditions:
$u(r_{ij} \rightarrow 0) \rightarrow -\infty$ and
$u(r_{ij} \rightarrow +\infty) \rightarrow 0$]. 
In the case of the BRS wave function, the correlation (pseudo) potential
is logarithmically long-range, however it is possible to extend the
method formally by splitting all quantities to compute into a short-
and long-range parts (see below).  It can be shown that the pair
distribution function can  be expressed as a series of cluster terms
associated with linked diagrams and will be given from the following
HNC equations: 
\bea
\label{composite}  
     X(\rr_{12}) &=& e^{ u(\rr_{12})+N(\rr_{12}) + 
                                 E(\rr_{12})  }-
                                 N(\rr_{12})-1 \,, \\
\label{nodal}  
  N(\rr_{12}) & = & \rho  \int d^2r_{3} \ X(\rr_{13}) \cdot
        [X(\rr_{32})+N(\rr_{32})] \,, \\
\label{g}  
  g(\rr_{12}) &=& 1+X(\rr_{12})+N(\rr_{12}) \,.
\eea
The quantities $X(\rr_{12})$  and $N(\rr_{12})$ represent the sum of
the so-called composite and nodal diagrams respectively and
$E(\rr_{12})$ is the sum of  elementary diagrams.  The generation of
diagrams contributing to $g(\rr_{12})$ must go through a
self-consistent procedure.  As a first approximation (and a good one)
we take the HNC/0 approximation where the ``0'' denotes the neglect of
elementary diagrams.  The summation of the nodal diagrams
$N(\rr_{12})$ is easily  performed in Fourier space.  

In order to handle the 2D logarithmic (pseudo) potential 
$u(\rr_{12})$, the standard procedure is to split it into short- and
long-range parts:
\begin{equation}
u(\rr_{12})= u_{s}(\rr_{12}) + u_{l}(\rr_{12}) \,,
\label{splitu}
\end{equation}
with the nodal function $N(\rr_{12})$ and the
composite function $X(\rr_{12})$ similarly split:
\bea
\label{split}
N(\rr_{12}) &=& N_{s}(\rr_{12}) - u_{l}(\rr_{12}) \,, \\
X(\rr_{12}) &=& X_{s}(\rr_{12}) + u_{l}(\rr_{12})  \,.
\eea
This splitting is done subject to the following conditions:
\bea
\label{splitcond}
u(\rr_{12}) + N(\rr_{12}) &=& u_s(\rr_{12}) + N_s(\rr_{12}) \,, \\
N(\rr_{12}) + X(\rr_{12}) &=& N_s(\rr_{12}) + X_s(\rr_{12}) \,.
\eea
Given the particular form of (pseudo) potential for the BRS wave function,
we choose to decompose $u(\rr_{12})$ into its short-range function
(going to -$\infty$ for small distances and healing to $0$ for large
distances) and its long-range counterpart in the following manner: 
\bea
\label{splitus}
u_{s}(\rr_{12}) &=& - 2 \, K_0(Q \, r_{12})  \\
	&& 
- 2 \, K_0(Q |\rr_{12}\!-\!\vec{\alpha}|) 
- 2 \, K_0(Q |\rr_{12}+\vec{\alpha}|)  \,,
\nn \\
u_{l}(\rr_{12}) &=&
  +	2 \, [ \ln(r_{12}) + K_0(Q r_{12})]  \\
\label{splitul}
&& +	2 \, [ \ln(|\rr_{12}-\vec{\alpha}|) + K_0(Q
|\rr_{12}-\vec{\alpha}|)] \nn \\
&& + 2 \, [ \ln(|\rr_{12}+\vec{\alpha}|)+ K_0(Q |\rr_{12}+\vec{\alpha}|)] 
\,, \nn
\eea
where $K_0(x)$ is the modified Bessel function, and $Q$ is a cut-off
parameter of order 1.  We recall that the 2D FT\cite{ft} of
$u_{l}(\rr_{12})$ is: 
\begin{equation}
\tilde{u}_{l}(\qq) = - \frac{4 \pi \, Q^2}{q^2(q^2+Q^2)} 
\left( 1 + e^{i \qq \cdot \vec{\alpha}} 
	+ e^{-i \qq \cdot \vec{\alpha}} \right) \,.
\label{ulq}
\end{equation}

The final set of equations is solved by initially setting
$N_s(\rr_{12})=0$ in Eq.\ (\ref{composite}), 
then obtaining  $X_{s}(\qq)= {\rm FT} [X_{s}(\rr_{12})]$ which can be
used to compute $\tilde{X}(\qq) = \tilde{X}_{s}(\qq) + \tilde{u}_l(\qq)$.
Using the convolution theorem we find 
$\tilde{N}(\qq) =\rho \tilde{X}(\qq)^2/[1 - \rho \tilde{X}(\qq)]$ and
easily obtain $\tilde{N}_s(\qq) = \tilde{N}(\qq) + \tilde{u}_l(\qq)$. 
The last step is to perform an inverse 2D FT on 
$\tilde{N}_s(\qq)$ to obtain the new  $N_s(\rr_{12})$.
This procedure is repeated until a desired accuracy is reached.
After convergence the pair distribution function is given by
\begin{equation}
\label{finalgr}
g(\rr_{12}) = 1 + X_s(\rr_{12}) + N_{s}(\rr_{12}) \,.
\ee
Simultaneously, the static structure factor is given by
\be
\label{finalsq}
	S(\qq) =
	1 + \rho \left[ \tilde{X}_s(\qq)+ \tilde{N}_s(\qq) \right] \,.
\ee
The computation of such functions allows us to find the interaction
energy and other related quantities.

\vspace{-0.3cm}
\section{Results and discussions}
\label{sec:results}
\vspace{-0.3cm}

In the present work we applied the HNC theory to study the BRS state
at filling $1/3$ of an arbitrary LL (in the single-LL approximation).
For the sake of simplicity we neglected the elementary diagrams (i.e.\
the so-called HNC/0 approximation).  This allows us to determine to a
reasonable accuracy the pair distribution function and the static
structure factor.  In order to compare the $\alpha=0$ (Laughlin)
state with the $\alpha \neq 0$ (BRS) state we studied the properties
of the BRS wave function for several $\alpha$-s with magnitudes
between $0$ to $3$ (in general 
$\alpha = |\alpha| \, e^{i \, \theta_\alpha}$, without loosing generality
we considered only  $\theta_\alpha = 0$).

\vspace{-0.3cm}
\subsection{Pair distribution function and structure factor}
\vspace{-0.3cm}

In Fig.\ \ref{fig:gr} we plot the pair distribution function $g(\rr)$
for $\alpha = 2$ (top, center panels), and the angle-averaged pair
distribution function $\overline{g}(r)$ corresponding to $\alpha$ = 0,
1, 2 and 3 (bottom panel). It is interesting to note the splitting of
the triple node at the origin, the noticeable angle-dependence of 
$g(\rr)$, and the change in the small-$r$ behavior of $\overline{g}(r)$
which switches from $\propto r^6$ (for $\alpha=0$) to $\propto r^2$ as
$\alpha$ is increased.  In addition, we mention the following generic
properties more or less valid for any of the $\alpha$-s we have
considered: 

\noindent
\begin{enumerate}
\tighten

\item
For $\alpha \neq 0$ there is at least one pair distribution
function that has an additional zero (besides the zero at $r=0$) at
inter-particle distance $r=\alpha$ and angle
$\theta=\theta_{\alpha}, \theta_\alpha+\pi$ ($\theta_\alpha=0$ in this case);  

\vspace{-0.15cm}
\item
For $\alpha \neq 0$ there are special inter-particle distances,
$r$ (besides the zero at the origin) where all pair distribution
functions cross, irrespective of their angle $\theta$ dependence.

\vspace{-0.15cm}
\item
Extremely interesting is the behavior of the angle-averaged pair
distribution function $\overline{g}(r)$ as a function of $\alpha$.
One notes that the major peak of $\overline{g}(r)$ simply shifts to
larger distances (without any sizeable change in its height) as
$\alpha$ is increased.  For smaller distances $\overline{g}(r)$ starts
to develop a shoulder that is quite visible for $\alpha=3$ contrary to
what seen in Ref.\ \onlinecite{joynt} for a slightly larger
$\alpha=3.2$, where the shoulder should had been even larger. 

\vspace{-0.15cm}
\item
For small-$r$, $g(\rr)$ has almost no angular dependence, and for
$\alpha \neq 0$, $g(r \approx 0,\theta) \simeq C_{\alpha} \, r^2$ for  
$0 \leq r \leq 0.5$, where $C_{\alpha} \simeq  0.026 \, \alpha^{2.5}$;
when $\alpha=0$, $g(r \approx 0) \propto \, r^6$ as expected.
These results derive immediately from the 1-fold vanishing of the BRS
wave function when two electrons come close, as opposed to the 3-fold
vanishing of Laughlin's wave function, since for small distances only
two-body correlations are important.
The absence of angular dependence on $g(r,\theta)$ for small $r$
can be easily understood by noting that in the small-$r$ limit:
$g(r \approx 0,\theta) \propto \exp[u_{s}(r \approx 0, \theta)]$, where
$ u_{s}(r, \theta)$ is given from Eq.\ (\ref{splitus}) and does not have
any angular dependence. By recalling that $\lim_{r \rightarrow 0}
K_{0}(Q r)=-\ln({Q r}/{2})-\gamma$, where $\gamma=0.5772...$ is
the Euler's constant one can easily understand why $g(r,\theta)$ has a
quadratic dependence on $r$ and not any angular dependence for small
$r$ and values of $\alpha \neq 0$. 
Although such quadratic dependence at short $r$ is also characteristic
for a Wigner crystal state at such filling factor \cite{smallr}
we note that the BRS state does not represent a crystalline state and
the overall pair distribution function of the BRS state is strikingly
different from the pair distribution function of the Wigner crystal
state.

\end{enumerate}


In Fig.\ \ref{fig:sq} we plot the static structure factor $S(\qq)$
for $\alpha = 2$ (top, center panels), and the angle-averaged
static structure factor $\overline{S}(q)$ corresponding to $\alpha$ =
0, 1, 2 and 3 (bottom panel).  The most important feature is the
emergence of peaks in $S(\qq)$ characteristic of a nematic structure. 
Broadly speaking, the major peak of the $\overline{S}(q)$
shifts to smaller $q$ and its height raises when $\alpha$ is
increased, with no significant change in the small-$q$
behavior. 

\vspace{-0.3cm}
\subsection{Energy of BRS states}
\vspace{-0.3cm}

One can compute the correlation energy per particle either directly
from Eqs.\ (\ref{potential}), (\ref{eq:ener_sq}), or
(\ref{eq:ener_sq_N}) to determine the energy per electron for
arbitrary values of the BRS parameter $\alpha$, the 2D system width
$\lambda$, and Landau level index $L$.  The following simplified
formula can be used in view of Eq.\ (\ref{angleav}):
\begin{equation}
E^L_{\alpha}(\lambda)= \frac{1}{4 \pi} \int_{0}^{\infty} \!\! dq \, q \,
\tilde{V}(q,\lambda) \, [L_L(\frac{q^2}{2})]^2 \, 
	[ \overline{S}(q) - 1 ] \,,
\label{enerfroms}
\end{equation}
where
$\tilde{V}(q,\lambda) = ({2 \pi \, e^2}/{\epsilon \, q}) 
\, \exp(-\lambda q)$ is the 2D FT of the ZDS interaction
potential.\cite{ZDS}  
In addition to allowing straightforward calculations to be extended to 
any LL, Eq.\ (\ref{enerfroms}) permits higher numerical accuracy on
the calculation of $E_\alpha$ since $\overline{S}(q)$ saturates
exponentially to $1$ for relatively small values of $q$
as compared to $\overline{g}(r)$.  

Figure \ref{fig:deltae} shows the energy difference between BRS states 
with $\alpha$ = 1, 2 and 3, and the isotropic state with $\alpha=0$.  
Our findings indicate that in the LLL ($L=0$) the Laughlin state is
stable for any $\lambda$, since all $\alpha \neq 0$ states have
higher energies (top panel), contrary to prior results \cite{joynt}
that the BRS state for $\alpha=1$ has a lower energy than the Laughlin
state if one considers the ZDS potential, with 
$\lambda > \lambda_c = 4.1 \pm 1.5$.

The situation changes drammatically in higher LLs ($L \ge 1$).  For
{\em small} $\lambda$ BRS states have {\em lower} energies and the
incompressible Laughlin-like state is unstable (see lower panels of
Fig.\ \ref{fig:deltae}).  The presence of this instability towards a
BRS state may be related to the absence of FQHE states in higher LLs,
since for $\alpha \neq 0$ highly damped low-energy modes exist in the
resulting nematic system. \cite{damping}  It is worth noting that for
$\lambda \lesssim 1$ the highest investigated $\alpha$ has the lowest
energy.  In this regime, we are therefore unable to determine the
optimal state (even within this familty of trial wavefunctions).
This BRS instability may be indicative of a transition towards 
a completely different state (e.g.\ as in Ref.\ \onlinecite{koulakov}).

At this point it is important to comment on how precise our
determination of these energy differences is.  The reader should note
that the HNC/0 approximation is essentially a variational method that
always gives an energy that constitutes an upper bound to the exact
ground state energy. \cite{ripka}  For example:  for the Laughlin
state with $\nu = 1/3$ and $\lambda = 0$, HNC/0 gives an adimensional
correlation energy of -0.405, while the ``exact'' value (determined by
Monte Carlo \cite{exact}) is -0.410.  Similar errors (ca.\ 1\%) will
be present for $\alpha \neq 0$ as well.  While an error of this
magnitude seems to be of the same order as, or bigger than, the energy
differences we are interested in, we remark that these are not {\em
uncorrelated errors} but {\em systematic deviations} due to the nature
of the approximations used, and energy {\em differences} are
considerably more precise.  Preliminary results using Monte Carlo
simulations \cite{mc} for a handful of cases indicate that energy
differences are, indeed, significant.



\vspace{-0.3cm}
\section{Conclusions, extensions and further developments}
\label{sec:conclusions}
\vspace{-0.3cm}


In conclusion, we applied the HNC theory to study possible BRS states
in a 1/3-filled LL.  We find that the isotropic Laughlin state is
stable in the LLL for realistic interaction potentials.  In higher
LLs, instabilities towards a BRS state are possible.  Since BRS states 
are gapless \cite{damping} this may be a simple explanation why no
FQHE was observed for 1/3-filled higher LLs.  One caveat is that the
magnitude of the energy differences obtained is comparable to the
absolute accuracy in the determination of individual energies by the
HNC method.  Although we believe that energy {\em differences} are
yielded more precisely than the energies themselves, these results
need confirmation by alternative (albeit more time-consuming) methods.
Monte Carlo simulations with large number of electrons are currently
being performed. \cite{mc} 

While these results are by themselves compelling, the connection to 
recent obsevations of liquid cristalline phases in half- and
quarter-filled LLs requires more sophisticated methods.  One
possibility is to generalize MJ's approach to composite fermion (CF) 
states, \cite{jain} e.g.\
\bea
\label{eq:MJhalf}
\Psi_\alpha^{1/2}(z_1, \ldots ,z_N) &=& \\
&& \hspace{-2.5cm}  P_{L} \,
\prod_{j<k} (z_j-z_k+\alpha)(z_j-z_k-\alpha) \; 
 {\rm Det} \left[ \phi_{\bf k}(\rr_i) \right]_{k < k_F} \,, \nn
\eea
where $\phi_{\bf k}(\rr_i)$ are plane waves of CFs, filled for
$k \! < \! k_F \! = \! (4 \pi \rho)^{1/2}$, and $P_{L}$ projects
onto the $L^{\rm th}$ LL.  This wavefunction is an obvious starting
point to study the {\em nematic} quantum Hall liquid crystals at half 
filling.  For the RIQHE observed near 1/4 filling, similar
generalizations are possible.  The presence of the Slater determinant
in the trial (in addition to the Jastrow factors) implies the need to
use the considerably more complex Fermi HNC \cite{ripka,orionfhnc}.
Calculations are under way for trial states of these form. \cite{half}

\vspace{-0.3cm}
\acknowledgments
\vspace{-0.3cm}

We would like to acknowledge helpful discussions with A.T.\ Dorsey,
M.\ Fogler and L.\ Radzihovsky. 
This work was supported by the University of Missouri Research Board.

\newpage
\section*{FIGURES}

\begin{figure}
\begin{center}
\leavevmode
\epsfxsize=3.4in
\epsfbox{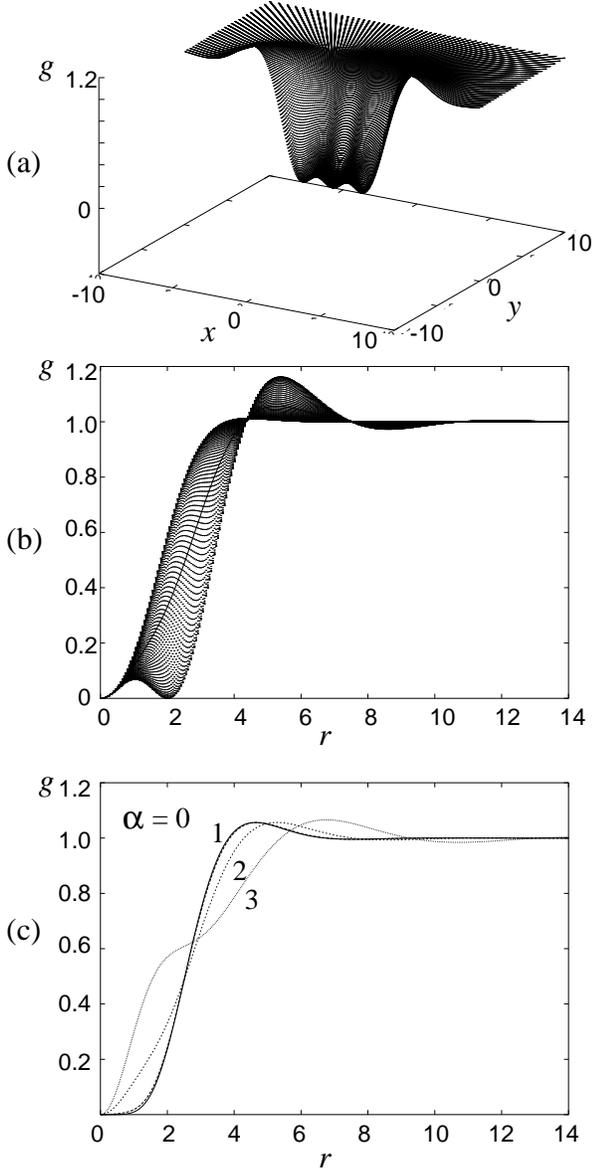}
\end{center}
\vspace{-0.cm}
\caption{ \label{fig:gr} Pair distribution function
for the BRS state at $\nu=1/3$. {\em (a)} $\alpha=2$, surface plot of
$g(r,\theta)$ (the surface for $y<0$ was removed for clarity); 
{\em (b)} $\alpha=2$, dotted lines: $g(r,\theta)$ for various 
$\theta \in [0,2\pi]$, full line: angle averaged $\overline{g}(r)$; 
{\em (c)} Angle averaged $\overline{g}(r)$ for 
$\alpha$ = 0, 1, 2 and 3 (0 and 1 are virtually identical).
Note the discrete nodes of $g(r,\theta)$ at $r = \alpha$, $\theta =
\theta_\alpha, \theta_\alpha + \pi$ ($\theta_\alpha = 0$ in this
case). Calculations were performed in the HNC/0 approximation.
}
\end{figure}

\newpage
\section*{}

\begin{figure}
\begin{center}
\leavevmode
\epsfxsize=3.4in
\epsfbox{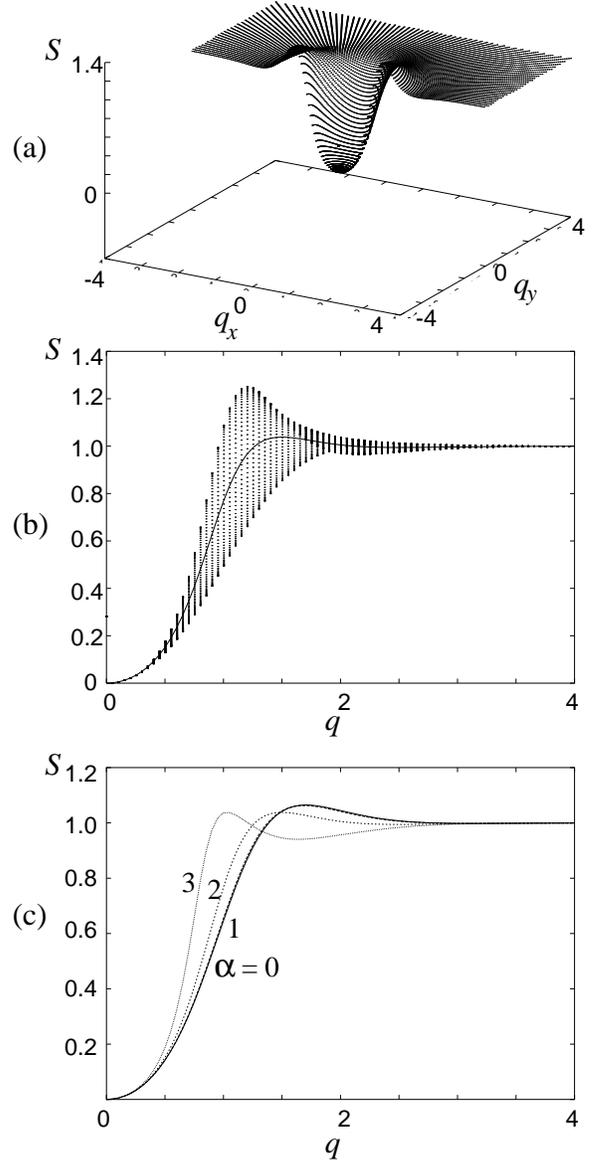}
\end{center}
\caption{ \label{fig:sq} Static structure factor 
for the BRS state at $\nu=1/3$. {\em (a)} $\alpha=2$, surface plot of
$S(r,\theta)$ (the surface for $q_y<0$ was removed for clarity); 
{\em (b)} $\alpha=2$, dotted lines: $S(q,\theta_q)$ for various 
$\theta_q \in [0,2\pi]$, full line: angle averaged $\overline{S}(q)$; 
{\em (c)} Angle averaged $\overline{S}(q)$ for 
$\alpha$ = 0, 1, 2 and 3 (0 and 1 are virtually identical).  Note the
presence of peaks in $S(\qq)$ consistent with a nematic
structure. Calculations were performed in the HNC/0 approximation. 
}
\end{figure}

\newpage
\widetext
\begin{figure}
\begin{center}
\leavevmode
\epsfxsize=6.6in
\epsfbox{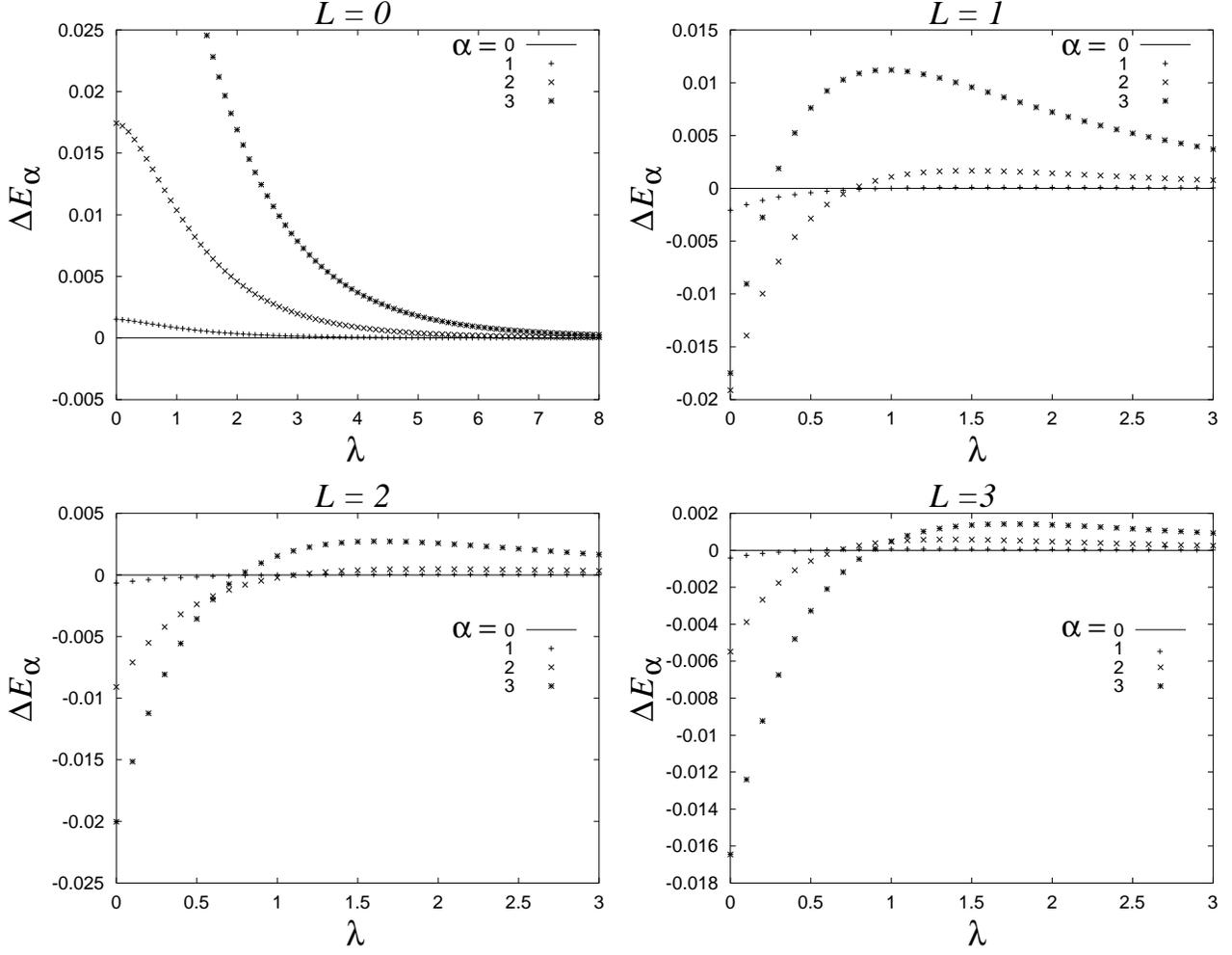}
\end{center}
\caption{  \label{fig:deltae}
	Energy per particle in BRS states with $\alpha$ = 1, 2 and 3
	relative to the isotropic ($\alpha=0$) state: 
	$\Delta E_{\alpha}(\lambda) = E_{\alpha}(\lambda) -
	E_{0}(\lambda)$ for various Landau levels $L$ as functions of
	the short distance cut-off $\lambda$ [Eq.\ (\protect\ref{enerfroms})].
	Energies are in units of $e^2/(\epsilon l_o)$.  Note that in
	the LLL ($L=0$) BRS states are always higher in energy,
	whereas in higher LLs ($L \ge 1$) there are ranges of
	$\lambda$ for which BRS states are favorable.
}
\end{figure}


\end{document}